\documentclass[fleqn,usenatbib]{mnras}
\hypersetup{draft} 

\usepackage{newtxtext,newtxmath}

\usepackage[T1]{fontenc}
\usepackage{ae,aecompl} 
\usepackage{lscape}

\usepackage{graphicx}	
\usepackage{amsmath}	
\usepackage{amssymb}	

\newcommand{\gaia}{{\it Gaia}}

\newcommand{\NGTS}{NGTS}

\newcommand{\LSO}{La Silla Observatory}

\newcommand{\Euler}{Eulercam}

\newcommand{\SAAO}{SAAO}
\newcommand{\SHOC}{SHOC}

\newcommand{\kms}{km\,s$^{-1}$}
\newcommand{\ms}{m\,s$^{-1}$}
\newcommand{\masy}{mas\,y$^{-1}$}
\newcommand{\mpl}{\mbox{M$_{p}$}}
\newcommand{\rpl}{\mbox{R$_{p}$}}
\newcommand{\mstar}{\mbox{M$_{s}$}}
\newcommand{\rstar}{\mbox{R$_{s}$}}
\newcommand{\mjup}{\mbox{M$_{J}$}}
\newcommand{\rjup}{\mbox{R$_{J}$}}
\newcommand{\msun}{\mbox{$M_{\odot}$}}
\newcommand{\rsun}{\mbox{$R_{\odot}$}}

\newcommand{\gccc}{g\,cm$^{-3}$}

\newcommand{\Nstar}{NGTS-8}			    							
\newcommand{\Nkamp}{\mbox{$149.95 \pm 3.56$}}

\newcommand{\NpropRA}{\mbox{$21.363\pm0.047$}}				            
\newcommand{\NpropDec}{\mbox{$-10.194\pm0.048$}}				        
\newcommand{\Nparallax}{\mbox{$2.3027\pm0.0299$}}				            
\newcommand{\Ngaia}{\mbox{$6840435777723109888$}}           			
\newcommand{\NGAIAmag}{$13.4954 \pm 0.0003$}							    			
\newcommand{\NGAIABPmag}{$13.9606 \pm 0.0015$}										
\newcommand{\NGAIARPmag}{$12.8780 \pm 0.0006$}										
\newcommand{\Ngaiadistance}{\mbox{$434.273\pm5.639$}}

\newcommand{\NstarradiusSPECIES}{\mbox{$0.98\pm0.02$}}                  
\newcommand{\NstarmassSPECIES}{\mbox{$0.89\,^{+0.05}_{-0.04}$}}         
\newcommand{\NteffSPECIES}{\mbox{$5241\pm50$}}                          
\newcommand{\NloggSPECIES}{\mbox{$4.41\pm0.03$}}                        
\newcommand{\NmetalSPECIES}{\mbox{$0.24\pm0.09$}}                       
\newcommand{\NvsiniSPECIES}{\mbox{$3.56\pm0.67$}}                       
\newcommand{\NvmacSPECIES}{\mbox{$1.49\pm0.64$}}                        
\newcommand{\NstarageSPECIES}{\mbox{$12.48\,^{+3.23}_{-3.68}$}}         

\newcommand{\NVmag}{$13.68 \pm 0.06$}									
\newcommand{\NBmag}{$14.59 \pm 0.03$}									
\newcommand{\Ngmag}{$14.12 \pm 0.03$}									
\newcommand{\Nrmag}{$13.43 \pm 0.06$}									
\newcommand{\Nimag}{$13.21 \pm 0.06$}									

\newcommand{\NRA}{\mbox{$21^{\rmn{h}} 55^{\rmn{m}} 54\fs2$}}            
\newcommand{\NDec}{\mbox{$-14\degr 04\arcmin 05\farcs 85$}}     		
\newcommand{\Ntwomass}{\mbox{$21555419$-$1404062$}}       	    		
\newcommand{\NJmag}{$12.14 \pm 0.02$}									
\newcommand{\NHmag}{$11.75 \pm 0.02$}									
\newcommand{\NKmag}{$11.64 \pm 0.02$}									

\newcommand{\NWmag}{$11.59 \pm 0.02$}									
\newcommand{\NWWmag}{$11.62 \pm 0.02$}									
\newcommand{\NWWWmag}{$11.86 \pm 0.38$}			    					

\newcommand{\Nplanet}{NGTS-8b}										
\newcommand{\Nperiod}{\mbox{$2.49970 \pm 0.00001$}}                     
\newcommand{\Nperiodshort}{\mbox{$2.49970$}}                            
\newcommand{\Nduration}{\mbox{$2.61 \pm 0.06$}}                         
\newcommand{\Ntc}{\mbox{$2457500.17830 \pm 0.00072$}}                   
\newcommand{\Necc}{\mbox{$0.010\,^{+0.014}_{-0.010}$}}                  
\newcommand{\Nmass}{\mbox{$0.93\,^{+0.04}_{-0.03}$}}                    
\newcommand{\Nradius}{\mbox{$1.09 \pm 0.03$}}                           
\newcommand{\Ndensity}{\mbox{$0.89\,^{+0.08}_{-0.07}$}}                          
\newcommand{\NTeq}{\mbox{$1345 \pm 19$}}                                
\newcommand{\Nrratio}{\mbox{$0.114 \pm 0.002$}}   
\newcommand{\Nau}{\mbox{$0.035 \pm 0.001$}}                             
\newcommand{\Naoverr}{\mbox{$7.60 \pm 0.18$}}                           
\newcommand{\Ninc}{\mbox{$86.9 \pm 0.5$}}

\newcommand{\Ncameraid}{811}                                            
\newcommand{\Nbaselinenights}{227}                                      
\newcommand{\Nbaselinestart}{the 21st of April 2016}                    
\newcommand{\Nbaselineend}{the 3rd of December 2016}                    
\newcommand{\Nnimages}{177\,799}                                        

\newcommand{\mask}{K0}                           

\newcommand{\NNstar}{NGTS-9}			    							
\newcommand{\NNkamp}{\mbox{$293.44 \pm 15.08$}}

\newcommand{\NNpropRA}{\mbox{$-6.078 \pm 0.057$}}				            
\newcommand{\NNpropDec}{\mbox{$1.723 \pm 0.063$}} 				            
\newcommand{\NNparallax}{\mbox{$1.6136 \pm 0.0416$}}				            
\newcommand{\NNgaia}{\mbox{$5678340222972504832$}}           		    	
\newcommand{\NNGAIAmag}{$12.6547 \pm 0.0002$}							    			
\newcommand{\NNGAIABPmag}{$12.9503 \pm 0.0015$}										    
\newcommand{\NNGAIARPmag}{$12.2157 \pm 0.0013$}										    
\newcommand{\NNgaiadistance}{\mbox{$619.732\pm15.977$}}

\newcommand{\NNstarradiusSPECIES}{\mbox{$1.38 \pm 0.04$}}                   
\newcommand{\NNstarmassSPECIES}{\mbox{$1.34 \pm 0.05$}}                     
\newcommand{\NNteffSPECIES}{\mbox{$6330 \pm 130$}}                          
\newcommand{\NNloggSPECIES}{\mbox{$4.37 \pm 0.20$}}                          
\newcommand{\NNmetalSPECIES}{\mbox{$0.31 \pm 0.15$}}                        
\newcommand{\NNvsiniSPECIES}{\mbox{$6.38 \pm 1.05$}}                       
\newcommand{\NNvmacSPECIES}{\mbox{$5.47 \pm 1.05$}}                        
\newcommand{\NNstarageSPECIES}{\mbox{$0.96 \pm 0.60$}}                       

\newcommand{\NNVmag}{$12.80 \pm 0.02$}									    
\newcommand{\NNBmag}{$13.36 \pm 0.04$}									    
\newcommand{\NNgmag}{$13.03 \pm 0.05$}									    
\newcommand{\NNrmag}{$12.65 \pm 0.02$}									    
\newcommand{\NNimag}{$12.55 \pm 0.07$}									    

\newcommand{\NNRA}{\mbox{$09^{\rmn{h}} 27^{\rmn{m}} 41\fs0$}}               
\newcommand{\NNDec}{\mbox{$-19\degr 20\arcmin 50\farcs 33$}}     		    
\newcommand{\NNtwomass}{\mbox{$09274096$-$1920515$}}       	    		    
\newcommand{\NNJmag}{$11.71 \pm 0.03$}									    
\newcommand{\NNHmag}{$11.49 \pm 0.02$}									    
\newcommand{\NNKmag}{$11.45 \pm 0.02$}									    

\newcommand{\NNWmag}{$11.39 \pm 0.02$}									    
\newcommand{\NNWWmag}{$11.42 \pm 0.02$}									    
\newcommand{\NNWWWmag}{$11.58 \pm 0.20$}			    					

\newcommand{\NNplanet}{NGTS-9b}										    
\newcommand{\NNperiod}{\mbox{$4.43527 \pm 0.00002$}}                        
\newcommand{\NNperiodshort}{\mbox{$4.43527$}}                               
\newcommand{\NNduration}{\mbox{$2.05 \pm 0.07$}}                            
\newcommand{\NNtc}{\mbox{$2457671.81086 \pm 0.00265$}}                      
\newcommand{\NNecc}{\mbox{$0.060^{+0.076}_{-0.052}$}}                     
\newcommand{\NNmass}{\mbox{$2.90 \pm 0.17$}}                       
\newcommand{\NNradius}{\mbox{$1.07 \pm 0.06$}}                              
\newcommand{\NNdensity}{\mbox{$2.93^{+0.53}_{-0.49}$}}                             
\newcommand{\NNTeq}{\mbox{$1448 \pm 36$}}                                   
\newcommand{\NNrratio}{\mbox{$0.080 \pm 0.004$}}   
\newcommand{\NNau}{\mbox{$0.058^{+0.003}_{-0.002}$}}                                
\newcommand{\NNaoverr}{\mbox{$9.06 \pm 0.31$}}                              

\newcommand{\NNinc}{\mbox{$84.1 \pm 0.4$}}

\newcommand{\NNcameraid}{806}                                           
\newcommand{\NNbaselinenights}{234}                                     
\newcommand{\NNbaselinestart}{the 8th of October 2016}                  
\newcommand{\NNbaselineend}{the 29th of May 2017}                       
\newcommand{\NNnimages}{167\,933}                                       

\newcommand{\Nmask}{G2}                           

\title[\Nplanet\ and \NNplanet]{\Nplanet\ and \NNplanet: two non-inflated hot-Jupiters}

\author[J. C. Costes et al.]{
\parbox{\textwidth}{
Jean C. Costes,$^{12}$\thanks{E-mail: \href{jcostes01@qub.ac.uk}{jcostes01@qub.ac.uk}}
Christopher~A.~Watson,$^{12}$
Claudia~Belardi,$^{5}$
Ian~P.~Braker,$^{5}$
Matthew~R.~Burleigh,$^{5}$
Sarah~L.~Casewell,$^{5}$
Philipp~Eigm\"uller,$^{6}$
Maximilian~N.~G{\"u}nther,$^{3,15,16}$ 
James~A.~G.~Jackman,$^{1, 2}$
Louise~D.~Nielsen,$^{4}$
Maritza~G.~Soto,$^{13}$
Oliver~Turner,$^{4}$ 
David~R.~Anderson,$^{1,2}$
Daniel~Bayliss,$^{1, 2}$
Fran\c{c}ois~Bouchy,$^{4}$
Joshua~T.~Briegal,$^{3}$
Edward~M.~Bryant,$^{1, 2}$
Juan~Cabrera,$^{6}$
Alexander~Chaushev,$^{10}$
Szilard~Csizmadia,$^{6}$
Anders~Erikson,$^{6}$
Samuel~Gill,$^{1,2}$
Edward~Gillen,$^{3}$
Michael~R.~Goad,$^{5}$
Matthew~J.~Hooton,$^{12}$
James~S.~Jenkins,$^{8, 9}$
James~McCormac,$^{1, 2}$
Maximiliano Moyano,$^{14}$
Didier~Queloz,$^{3}$
Heike Rauer,$^{6,10,11}$
Liam~Raynard,$^{5}$
Alexis~M.~S.~Smith,$^{6}$
Andrew~P.~G.~Thompson,$^{12}$
Rosanna~H.~Tilbrook,$^{5}$
Stephane~Udry,$^{4}$
Jose I. Vines,$^{8}$
Richard~G.~West,$^{1, 2}$
Peter~J.~Wheatley$^{1,2}$
}
\vspace{0.3cm}
\\
$^{1}$Centre for Exoplanets and Habitability, University of Warwick, Gibbet Hill Road, Coventry CV4 7AL, UK\\
$^{2}$Dept.\ of Physics, University of Warwick, Gibbet Hill Road, Coventry CV4 7AL, UK\\
$^{3}$Astrophysics Group, Cavendish Laboratory, J.J. Thomson Avenue, Cambridge CB3 0HE, UK\\
$^{4}$Observatoire de Gen{\`e}ve, Universit{\'e} de Gen{\`e}ve, 51 Ch. des Maillettes, 1290 Sauverny, Switzerland\\
$^{5}$Department of Physics and Astronomy, University of Leicester, University Road, Leicester LE1 7RH, UK\\
$^{6}$Institute of Planetary Research, German Aerospace Center, Rutherfordstrasse 2, 12489 Berlin, Germany\\
$^{7}$Institute of Astronomy, Cambridge University, Madingley Road, Cambridge CB3 0HA, UK\\
$^{8}$Departamento de Astronomia, Universidad de Chile, Casilla 36-D, Santiago, Chile\\
$^{9}$ Centro de Astrof\'isica y Tecnolog\'ias Afines (CATA), Casilla 36-D, Santiago, Chile.\\
$^{10}$Center for Astronomy and Astrophysics, TU Berlin, Hardenbergstr. 36, D-10623 Berlin, Germany\\
$^{11}$Institute of Geological Sciences, FU Berlin, Malteserstr. 74-100, D-12249 Berlin, Germany\\
$^{12}$Astrophysics Research Centre, School of Mathematics and Physics, Queen's University Belfast, BT7 1NN Belfast, UK\\
$^{13}$School of Physics and Astronomy, Queen Mary University of London, 327 Mile End Road, E1 4NS, UK\\
$^{14}$Instituto de Astronom\'{i}a, Universidad Cat\'{o}lica del Norte, Angamos 0610, 1270709 Antofagasta, Chile\\
$^{15}$Department of Physics, and Kavli Institute for Astrophysics and Space Research,\\ Massachusetts Institute of Technology, Cambridge, MA 02139, USA\\
$^{16}$Juan Carlos Torres Fellow
}
\date{Accepted 2019 Nov 04. Received 2019 Oct 08; in original form 2019 Aug 05}

\pubyear{2019}

\begin{document}
\label{firstpage}
\pagerange{\pageref{firstpage}--\pageref{lastpage}}
\maketitle

\begin{abstract}

We report the discovery, by the Next Generation Transit Survey (NGTS), of two hot-Jupiters \Nplanet\ and \NNplanet. These orbit a $V\,=\,13.68$ K0V star ($T_{eff}$\,=\,\NteffSPECIES\,K) with a period of $\Nperiodshort$ days, and a $V\,=\,12.80$ F8V star ($T_{eff}$\,=\,\NNteffSPECIES\,K) in $\NNperiodshort$ days, respectively. The transits were independently verified by follow-up photometric observations with the SAAO 1.0-m and Euler telescopes, and we report on the planetary parameters using HARPS, FEROS and CORALIE radial velocities. \Nplanet\ has a mass, \Nmass\,\mjup\ and a radius, \Nradius\,\rjup\ similar to Jupiter, resulting in a density of \Ndensity\,\gccc. This is in contrast to \NNplanet, which has a mass of \NNmass\,\mjup\ and a radius of \NNradius\,\rjup, resulting in a much greater density of \NNdensity\,\gccc. Statistically, the planetary parameters put both objects in the regime where they would be expected to exhibit larger than predicted radii. However, we find that their radii are in agreement with predictions by theoretical non-inflated models.

\end{abstract}



\section{Introduction}
\label{sec:intro}

Hot-Jupiters are giant gas exoplanets similar to Jupiter, but with a shorter orbital period, inferior to 10\,days. While rare, these planets are the easiest to detect from ground-based surveys due to their relatively deep transits ($\sim 1\%$), their large radial velocity (RV) signals, and their short periods, which make hot-Jupiters important targets in order to understand the structure, composition and evolution of planetary systems.

From the currently observed population of exoplanets with known radii, masses and orbital distances, the evolution of planetary radii has been modelled (e.g. \citealt{Fortney2007}; \citealt{Baraffe2008}). These models, where the effects of stellar irradiation and heavy element cores are included, agree with observations at low stellar irradiation. However, the observed radii of highly irradiated gas giants are discrepant with theoretical expectations. For instance, at fluxes greater than $2 \times 10^5$\,W\,m$^{-2}$ (\citealt{Miller2011}; \citealt{Demory2011}) the gas giants are increasingly found with anomalously large radii. This is the case for the hot-Jupiters WASP-17 b, WASP-121 b and Kepler-435 b, which all have measured radii R\,>\,1.8\,\rjup (\citealt{Anderson2011}; \citealt{Almenara2015}; \citealt{Delrez2016}). A number of possible mechanisms have been postulated to  explain these inflated planetary radii including kinetic heating \citep{Guillot2002}, enhanced atmospheric opacities \citep{Burrows2007}, double diffusive convection \citep{Chabrier2007}, Ohmic heating through magnetohydrodynamic effects (\citealt{Batygin2010}; \citealt{Perna2010}; \citealt{Wu2012}; \citealt{Ginzburg2016}), tidal dissipation (\citealt{Bodenheimer2001}; \citealt{Bodenheimer2003}; \citealt{Arras2010}; \citealt{Jermyn2017}) and vertical advection of potential temperature (\citealt{Youdin2010}; \citealt{Tremblin2017}). However, the exact mechanisms responsible are still, as yet, unidentified and the problem remains unsolved. In order to perform robust statistical studies of hot-Jupiter radii and constrain the dominant `inflation' mechanisms at work (e.g. as done by \citealt{Sestovic2018}) we need to increase the sample of planets spanning a range of planetary masses, radii, stellar irradiation levels, as well as planetary system ages.

In this paper we present two hot-Jupiters that appear to be non-inflated, despite being highly irradiated with an incident flux greater than $2 \times 10^5$\,W\,m$^{-2}$ (like many inflated planets). In \S\ref{sec:ngts}, the NGTS discovery data is described. \S\ref{sec:phfu} explains the photometric follow-up campaigns and \S\ref{sec:spec} reports the mass determination via RV monitoring from spectroscopy. \S\ref{sec:analysis} details the analysis of the stellar parameters, presents the stellar activity and its relation with the stellar rotation and shows the global modelling process to characterize the planets. \S\ref{sec:Discussion} presents an investigation regarding the incident flux, the planetary mass and the radius. Finally we finish with our conclusions in \S\ref{sec:conclusions}.


\section{Discovery Photometry From \NGTS}
\label{sec:ngts}

The Next Generation Transit Survey (NGTS), operating since early 2016, is a wide-field transit survey located at ESO's Paranal Observatory in Chile, whose primary goal is to discover Neptune-sized or bigger exoplanets. NGTS has a fully robotized array of twelve 20 cm Newtonian telescopes, and each telescope is equipped with 2K$\times$2K e2V deep-depleted Andor IKon-L CCD cameras with 13.5\,\textmu{}m pixels and an instantaneous field of view of 8 deg$^{2}$. For a description of this facility and its capabilities, optimised for detecting planets, we refer the reader to \cite{Wheatley2018}. NGTS has already detected 4 hot-Jupiters: NGTS-1b \citep{Bayliss2018}, NGTS-2b \citep{Raynard2018}, NGTS-3Ab \citep{Gunther2018} and NGTS-6b \citep{Vines2019}. Here we report the latest hot-Jupiter discoveries from NGTS: \Nplanet\ and \NNplanet.

\Nstar\ was observed using a single NGTS camera (\#\Ncameraid) over a \Nbaselinenights~night baseline between \Nbaselinestart~and \Nbaselineend. \NNstar~ was also observed using a single camera (\#\NNcameraid) over a \NNbaselinenights~night baseline between \NNbaselinestart~and \NNbaselineend. A total of \Nnimages\ and \NNnimages\ images were obtained, respectively, each with an exposure time of $10$\,s. These data were taken using the custom \NGTS~filter (520 -- 890\,nm) \citep{Wheatley2018} and the telescope was auto-guided using an improved version of the DONUTS auto-guiding algorithm \citep{McCormac2013}. The data were reduced and aperture photometry was extracted using the CASUTools\footnote{\url{http://casu.ast.cam.ac.uk/surveys-projects/software-release}} photometry package. A total of 177120 and 166043 valid data-points were extracted from the raw images and then de-trended for nightly trends, such as atmospheric extinction, using our implementation of the SysRem algorithm \citep{Tamuz2005}.

\begin{figure}
	\centering
	\includegraphics[width=0.855\columnwidth]{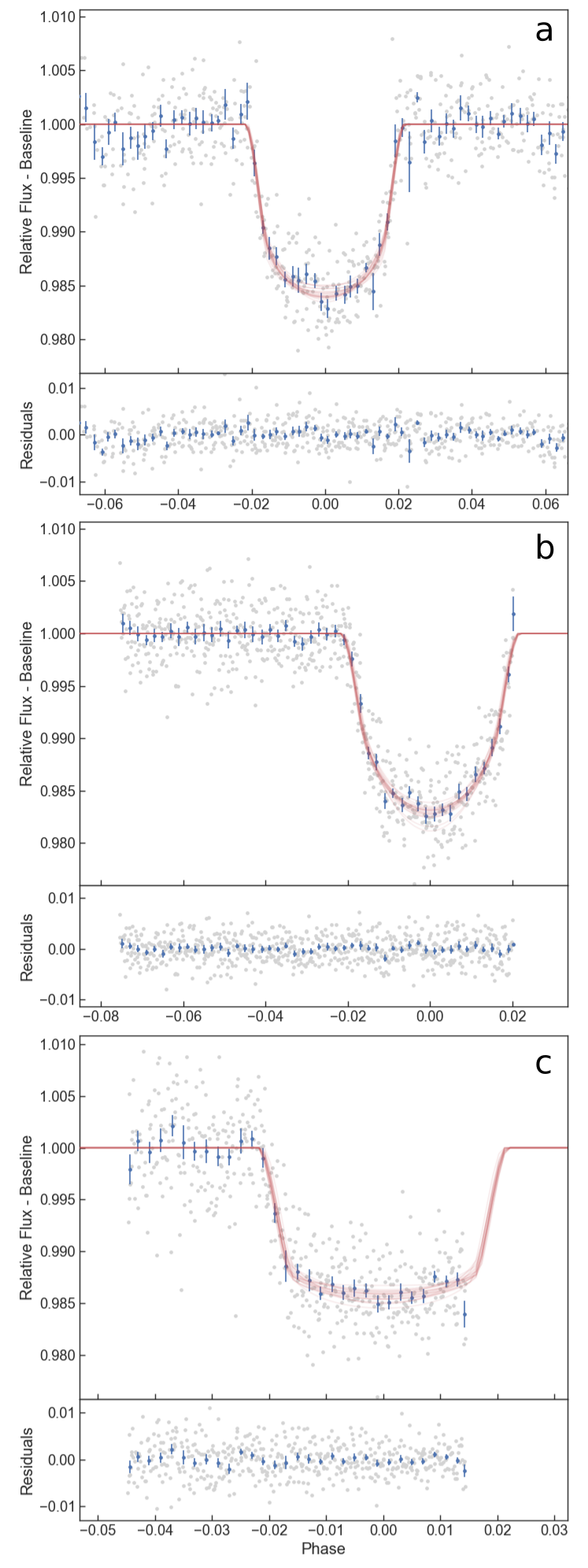}
    \caption{From top to bottom: Figure 1.a represents the \NGTS\ discovery light curve of \Nplanet\ with residuals, phase-folded on the orbital period and zoomed on the transit. Figure 1.b represnts the ingress and mid-transit of \Nplanet\ observed with SAAO, with residuals. Figure 1.c represents the ingress and mid transit of \Nplanet\ observed with Euler, with residuals. For all the plots, the blue data points are binned every 7\,min to aid visualisation. The red lines show 20 light curve models generated from randomly drawn posterior samples of the \texttt{allesfitter} fit.} 
    \label{fig:ngtsphot1}
\end{figure}

\begin{figure}
	\centering
	\includegraphics[width=0.85\columnwidth]{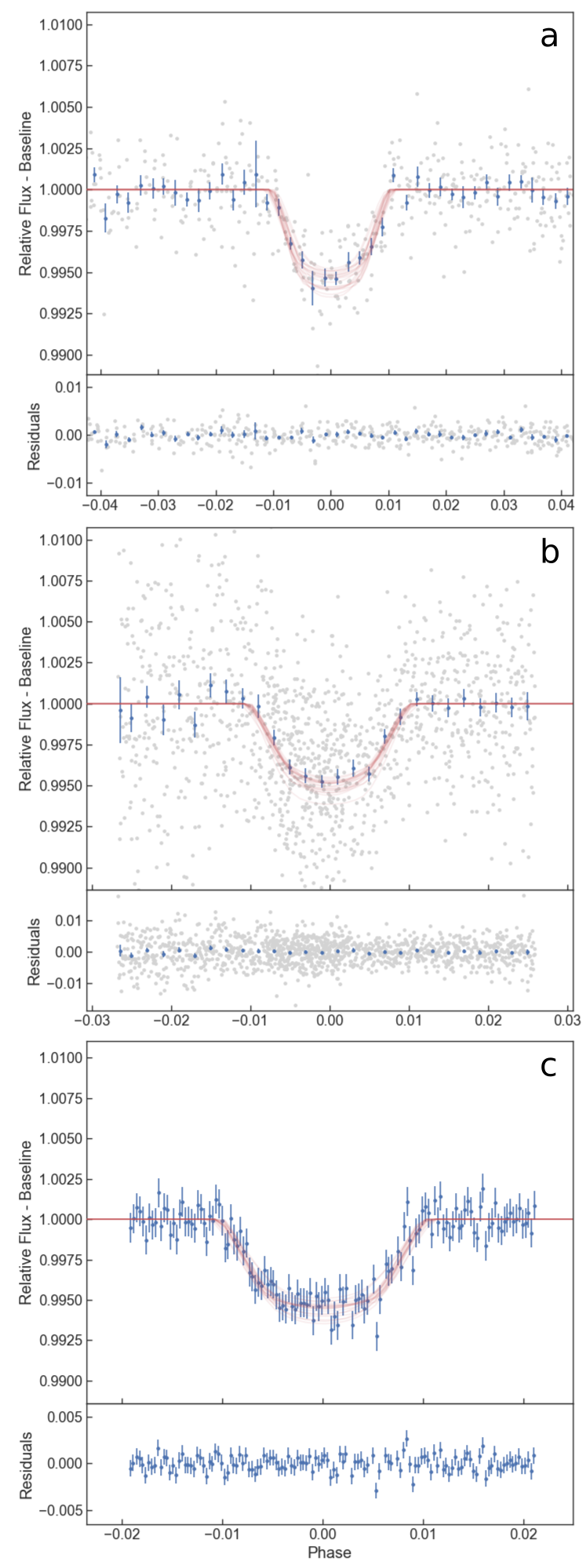}
    \caption{From top to bottom: Figure 2.a represents the \NGTS\ discovery light curve of \NNplanet\ with residuals, phase-folded on the orbital period and zoomed on the transit. Figure 2.b represents the full transit of \NNplanet\ observed with SAAO, with residuals. This image uses night 20181221, where ingress and mid-transit can be seen, and night 20190103, with mid-transit and egress. For both figures, the blue data points are binned every 10\,min to aid visualisation. Figure 2.c represents the full transit of \NNplanet\ observed with Euler, with residuals. The data was taken on a 100sec cadence, shown in blue. For all the plots, the red lines show 20 light curve models generated from randomly drawn posterior samples of the \texttt{allesfitter} fit.} 
    \label{fig:ngtsphot2}
\end{figure}

Both datasets were searched for transit-like signals using {\sc orion}, an optimized implementation of the box-least-squares (BLS) fitting algorithm \citep{colliercameron2006}. A 1.6\% deep transit signal was detected at a period of \Nperiodshort\,days for the K0V star, \Nstar, and a 0.6\% deep transit signal at \NNperiodshort\,days for the F8V star, \NNstar. These periods were distinguished from other aliases using the photometry and spectroscopy follow-up -- see Section~\ref{sec:phfu}  and Section~\ref{sec:spec} for details. The detrended \NGTS{} data for the two stars, phase-folded on the planetary orbital periods, are shown in Figure~\ref{fig:ngtsphot1} and Figure~\ref{fig:ngtsphot2}. A sample of the NGTS reduced photometric measurements are presented in Table~\ref{tab:allphot1} and Table~\ref{tab:allphot2}, with the full data available electronically from the journal. 

The NGTS data were searched for signs that would indicate that the planetary candidates were false positives. No evidence for a secondary eclipse or out-of-transit variations indicating an eclipsing binary system were identified in the NGTS light curves of the two stars. However, for both sources, some stars were found to be in close proximity to our targets. Using \gaia\, we confirmed that these nearby stars did not appreciably dilute the light from \Nstar\ or \NNstar, and also confirmed that \Nstar\ and \NNstar\ were not giants stars -- see Section~\ref{secGaia} for details. Based on the \NGTS\ detection, \Nstar\ and \NNstar\ were followed-up with further photometry and spectroscopy to confirm the planetary nature of the system and to measure the planetary parameters, which we report on in the next section.

\begin{table}
	\centering
	\caption{A sample of the photometric data of \Nstar{} from \NGTS, \SAAO\ and \Euler. The full dataset is available electronically from the journal.}
	\label{tab:allphot1}
	\begin{tabular}{ccccc}
	Time    &   Relative    &   Flux    &   Filter  &   Instrument\\
    (BJD-2450000)   &   flux    &   error   &   &   \\
	\hline
    7499.8655	&	1.0179	&	0.0221	&	NGTS	&	NGTS	\\
    7499.8657	&	1.0320	&	0.0221	&	NGTS	&	NGTS	\\
    7499.8658	&	1.0016	&	0.0220	&	NGTS	&	NGTS	\\
    7499.8660	&	0.9964	&	0.0220	&	NGTS	&	NGTS	\\
    7499.8661	&	0.9911	&	0.0220	&	NGTS	&	NGTS	\\
    7499.8663	&	0.9995	&	0.0220	&	NGTS	&	NGTS	\\
    7499.8664	&	0.9892	&	0.0220	&	NGTS	&	NGTS	\\
    7499.8666	&	1.0054	&	0.0220	&	NGTS	&	NGTS	\\
    7499.8667	&	0.9910	&	0.0219	&	NGTS	&	NGTS	\\
    7499.8669	&	0.9901	&	0.0219	&	NGTS	&	NGTS	\\
	...	&	...	&	...	&	...	&	...\\
	\hline
	\end{tabular}
\end{table}

\begin{table}
	\centering
	\caption{A sample of the photometric data of \NNstar{} from \NGTS, \SAAO\ and \Euler. The full dataset is available electronically from the journal.}
	\label{tab:allphot2}
	\begin{tabular}{ccccc}
	Time    &   Relative    &   Flux    &   Filter  &   Instrument\\
    (BJD-2450000)   &   flux    &   error   &   &   \\
	\hline
    7669.8612	&	0.9871	&	0.0087	&	NGTS	&	NGTS	\\
    7669.8614	&	1.0027	&	0.0087	&	NGTS	&	NGTS	\\
    7669.8617	&	0.9919	&	0.0087	&	NGTS	&	NGTS	\\
    7669.8618	&	0.9848	&	0.0087	&	NGTS	&	NGTS	\\
    7669.8620	&	1.0107	&	0.0087	&	NGTS	&	NGTS	\\
    7669.8621	&	1.0014	&	0.0087	&	NGTS	&	NGTS	\\
    7669.8623	&	0.9956	&	0.0087	&	NGTS	&	NGTS	\\
    7669.8624	&	1.0167	&	0.0087	&	NGTS	&	NGTS	\\
    7669.8626	&	1.0076	&	0.0087	&	NGTS	&	NGTS	\\
    7669.8627	&	1.0000	&	0.0087	&	NGTS	&	NGTS	\\
	...	&	...	&	...	&	...	&	...\\
	\hline
	\end{tabular}
\end{table}

\begin{table*}
	\centering
	\caption{A summary of the follow-up photometry of \Nstar\ and \NNstar.}
	\label{tab:photsummary}
	\begin{tabular}{ccccccccc}
    Night & Instrument & Target &  N$_{\mathrm{images}}$ & 
    Exptime & Binning & Filter  & Comment \\
        &  &  & & (seconds) & (X$\times$Y) &  &  \\
	\hline
    20170717 & Shoc'n'awe & \Nplanet & 676 & 30 & $4\times4$ & z' &   shown in Figure~\ref{fig:ngtsphot1}.b \\
    20170718 & Shoc'n'awe & \Nplanet & 606 & 60 & $4\times4$ & z' &  no transit observed \\
    20170821 & Eulercam & \Nplanet &    502 & 12 & $1\times1$ & IC  &   shown in Figure~\ref{fig:ngtsphot1}.c \\
    20181221 & Shoc'n'awe & \NNplanet & 550 & 20 & $4\times4$ & $I$    & shown in Figure~\ref{fig:ngtsphot2}.b \\
    20190103 & Shoc'n'awe & \NNplanet & 648 & 20 & $4\times4$ & $I$    & shown in Figure~\ref{fig:ngtsphot2}.b  \\
    20190112 & Eulercam & \NNplanet &    134 & 100 & $1\times1$ & RG &   shown in Figure~\ref{fig:ngtsphot2}.c \\
    \hline
	\end{tabular}
\end{table*}


\section{Follow-up Photometry}
\label{sec:phfu}

\subsection{SAAO photometry}
\label{sub:saaophot}

Follow-up photometry of \Nstar\ was obtained with the 1.0\,m Elizabeth telescope at the South African Astronomical Observatory (\SAAO) on 2017 July 17 and 2017 July 18, utilising the frame-transfer CCD Sutherland High-speed Optical Camera "{\em SHOC'n'awe}" \citep[SHOC]{Coppejans2013}. 

With a pixel scale of 0.167 arcsec/pixel, the \SHOC~cameras on the 1\,m telescope were binned $4\times4$ pixels in the X and Y directions. The field of view of its instruments is $2.85\arcmin \times 2.85$\arcmin. This allow to observe simultaneously the target and a comparison star of similar brightness for differential photometry. The data, obtained using a $z'$ filter with an exposure of 30\,s, were bias and flat field corrected. This was performed in python using the standard procedure with the CCDPROC package \citep{Craig2015}. Then, using the `SEP' package \citep{Barbary2016}, the aperture photometry of both the target and the comparison star were extracted. Finally, the sky background was measured and substracted using the SEP background map.

We also obtained two follow-up light curves of \NNstar\ on 2018 December 21 and 2019 January 30, with the same telescope and instrument set-up as described above. This time the data were obtained with an $I$ filter and an exposure time of 20\,s. The data were reduced with the local SAAO SHOC pipeline, which is driven by {\sc python} scripts running {\sc iraf} tasks ({\sc pyfits} and {\sc pyraf}), and incorporating the usual bias and flat-field calibrations. Aperture photometry was performed using the Starlink package {\sc autophotom}. For the first observation of \NNstar\ we used a 4 pixel radius aperture that maximised the signal/noise, and the background was measured in an annulus surrounding this aperture with inner and outer radii of 7 and 9 pixels, respectively. Two comparison stars were then used to perform differential photometry on the target. The 2019 January 30 observation was obtained in slightly poorer seeing conditions, and we therefore utilised a 6 pixel aperture, a correspondingly larger background annulus, and only one comparison star for differential photometry. 

The transits of these two exoplanets observed from SAAO are shown in Figures~\ref{fig:ngtsphot1}.b and~\ref{fig:ngtsphot2}.b. Regarding \Nstar, only a partial transit was observed with SAAO. For \NNstar, 2 nights were combined: night 20181221, where ingress and mid-transit were seen, and night 20190103, where mid-transit and egress were seen. While only partial transits were observed, these SAAO data were able to confirm the transits and the consistency of the transit depths and were used to revise the orbital ephemerides for subsequent follow-up observations. In particular, the 2017 July observations of \Nstar\ were helpful in confirming the orbital period and ruling out aliases of similar power in the original NGTS data.

\subsection{Eulercam}
\label{sub:eulercam}

We also observed both objects with \Euler~\citep{Lendl2012} on the 1.2\,m Euler Telescope at \LSO. \Nstar{} was observed on the 21st of August 2017, 502 exposures were acquired using the Cousins-I filter, an exposure time of 12\,s and a defocus of 0.05\,mm. \NNstar{} was observed on the 12th of January 2019. We acquired 134 images using the Gunn-R filter, a 100\,s exposure time and no defocus. For both target, their data were reduce using the standard procedure of bias subtraction and flat field correction. The aperture photometry as well as x- and y-position, FWHM, airmass and sky background of the target star were extracted using the PyRAF implementation of the phot routine. The comparison stars and the photometric aperture radius were carefully chosen in order to reduce the RMS in the scatter out of transit.

Using both follow-up photometry for the 2 stars, we can conclude that the nearby stars did not blend with the two targets. The Euler data for the two stars are shown in Figures~\ref{fig:ngtsphot1}.c and~\ref{fig:ngtsphot2}.c. Regarding \Nstar, only a partial transit was observed with Euler. Concerning \NNstar, the full transit was observed, with some systematics that were removed using a Gaussian process. Again, these data confirm the transits as we see the ingress or the full transit around the predicted times.


\section{Spectroscopy}
\label{sec:spec}

\subsection{\Nstar}
\Nstar\ was observed with the HARPS spectrograph \citep{Mayor2003} on the ESO 3.6\,m telescope at \LSO, Chile, between the 5th of August 2017 and the 28th of October 2017 under programmes 099.C-0303 and 0100.C-0474. We used the high-efficiency mode, EGGS, due to the faintness of the host star and large expected RV amplitude. The exposure times for each spectrum varied between 1800 and 1200\,s resulting in a signal-to-noise (SNR), measured around 550\,nm, of 10-15 per exposure. The standard HARPS data reduction software (DRS) was used to measure the RVs of \Nstar{} at each epoch. This was done via cross-correlation with a \mask ~binary mask. 

Three additional spectra were obtained with FEROS \citep{KauferPasquini1998}, mounted on the MPG 2.2\,m telescope at \LSO, Chile, on the 20th and 21st of August 2017. All spectra were obtained with an exposure time of 1800\,s, and the data were reduced using the FEROS routine of the CERES pipeline \citep{Brahm2017}. CERES also performed a radial velocity extraction, by cross-correlating the spectra with a G2 binary mask. The resulting SNR of the spectra, taken around 550\,nm per resolution element, was around 45.

\begin{figure}
	\includegraphics[width=\columnwidth]{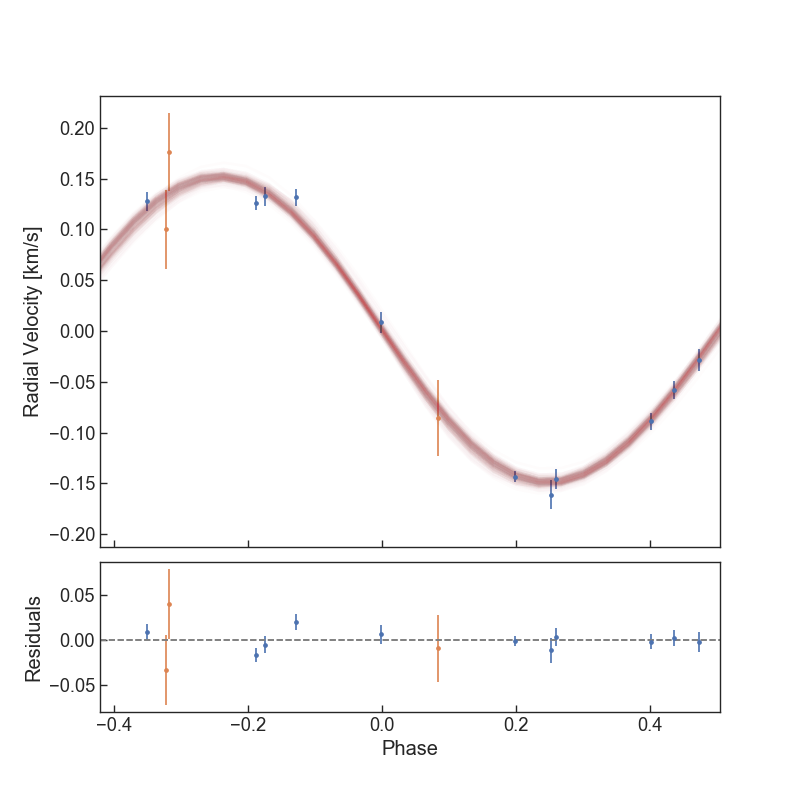}
    \caption{Phase folded radial velocity data in \kms\ and residuals from HARPS, in blue, and FEROS, in orange, for \Nstar. The red lines show 50 light curve models generated from randomly drawn posterior samples of the \texttt{allesfitter} fit.}
    \label{fig:harps1}
\end{figure}

The RVs from both HARPS and FEROS are listed in Table~\ref{tab:rvs1}, along with their associated error, FWHM and bisector span. While not presented in this table, the error on the BIS and on the FWHM were calculated using the same standard treatment as done previously \citep{West2018}. The errors for the BIS and for the FWHM are set to equal twice the error and 2.3548 times the error on the RV, respectively. The RV measurements of \Nstar, shown phase folded in Figure~\ref{fig:harps1}, are in phase with the period detected by {\sc orion}, with a semi-amplitude of $K=$\,\Nkamp\,\ms. This indicates a transiting planet with the mass of a hot-Jupiter. No evidence of a correlation between the RVs and the measured bisector spans or FWHMs were found, with a Spearman correlation of -0.05 and -0.21, respectively. Thus, the RV signal does not originate from cool stellar spots or a blended eclipsing binary \citep{queloz2001}.

\subsection{\NNstar}
\begin{figure}
	\includegraphics[width=\columnwidth]{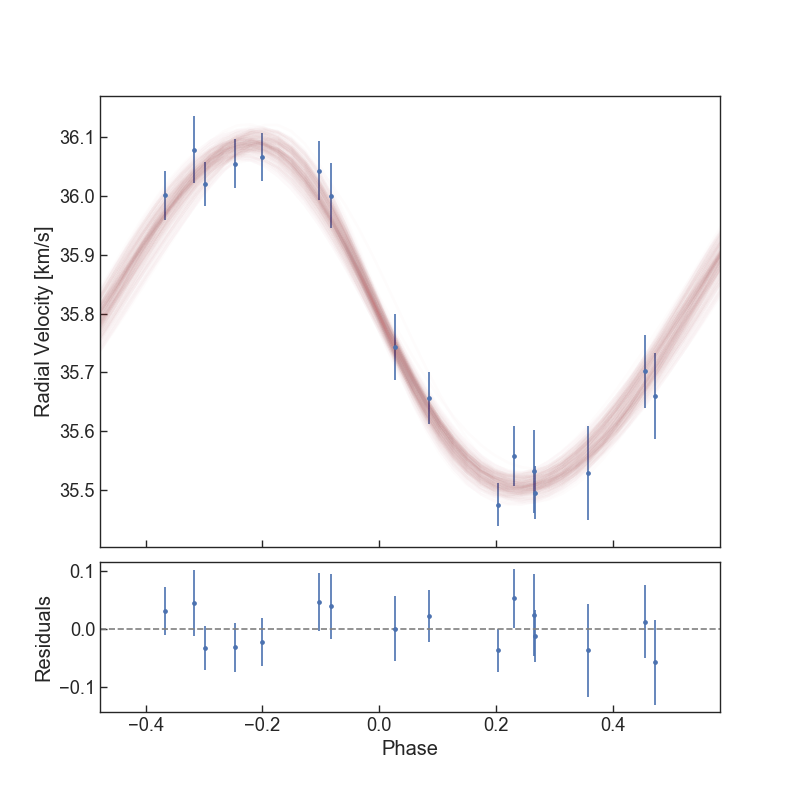}
    \caption{Phase folded radial velocity data in \kms\ and residuals from CORALIE for \NNstar. The red lines show 50 light curve models generated from randomly drawn posterior samples of the \texttt{allesfitter} fit.}
    \label{fig:harps2}
\end{figure}

\NNstar\ was observed with the CORALIE spectrograph \citep{CORALIE} on the 1.2\,m Euler telescope at \LSO, Chile, between the 24th of December 2017 and the 5th of April 2018. Exposure times of either 1800\, or 2700\,s were used depending on seeing and general observing conditions at the time. 
RVs were calculated with a \Nmask ~binary mask using the standard data reduction pipelines. Initial analysis, shown phase folded in Figure~\ref{fig:harps2}, confirm the planetary nature of the candidate with a mass of a hot-Jupiter. The RV variations are in phase with the period detected by {\sc orion} with a semi-amplitude of $K=$\,\NNkamp\,\ms. Correlations between RVs and bisector span and FWHM were also checked but no evidence for any such correlations was found, with a Spearman correlation of -0.05 and -0.15, respectively. The CORALIE RVs for \NNstar\ are listed in Table~\ref{tab:rvs2}, along with their associated error, FWHM and bisector span.

\begin{table*}
	\centering
	\caption{HARPS and FEROS radial velocities for \Nstar.}
	\label{tab:rvs1}
	\begin{tabular}{ccccccc} 
		JDB &   RV  &   RV error    &   FWHM    &   BIS &   Exptime &   Instrument\\
		(-2400000)	& (\kms)& (\kms) &(\kms) & (\kms) &  (s)    &  \\
		\hline
57970.7400	&	15.4538	&	0.0141	&	7.0583	&	-0.1398	&	1800	&	HARPS\\
57979.7893	&	15.7466	&	0.0084	&	6.9506	&	-0.0254	&	1800	&	HARPS\\
57980.7591	&	15.4692	&	0.0096	&	6.9007	&	-0.0021	&	1800	&	HARPS\\
57986.8002	&	15.3768	&	0.0147	&	10.4776	&	0.0000	&	1800	&	FEROS\\
57986.8147	&	15.4527	&	0.0147	&	10.4080	&	0.0360	&	1800	&	FEROS\\
57987.8182	&	15.1913	&	0.0108	&	10.5177	&	-0.0620	&	1800	&	FEROS\\
57993.6958	&	15.5568	&	0.0089	&	6.9108	&	0.0092	&	1800	&	HARPS\\
57994.6689	&	15.7472	&	0.0093	&	6.9062	&	-0.0238	&	1800	&	HARPS\\
57998.7907	&	15.5861	&	0.0109	&	6.9314	&	-0.0221	&	1200	&	HARPS\\
58023.6064	&	15.5259	&	0.0082	&	6.9028	&	-0.0128	&	1800	&	HARPS\\
58025.5985	&	15.4713	&	0.0057	&	6.9379	&	-0.0243	&	1800	&	HARPS\\
58026.7251	&	15.7423	&	0.0091	&	6.8863	&	-0.0141	&	1800	&	HARPS\\
58052.5954	&	15.6232	&	0.0107	&	6.9310	&	-0.0558	&	1200	&	HARPS\\
58054.6266	&	15.7407	&	0.0073	&	6.9014	&	-0.0074	&	1200	&	HARPS\\
		\hline
	\end{tabular}
\end{table*}

\begin{table*}
	\centering
	\caption{CORALIE radial velocities for \NNstar.}
	\label{tab:rvs2}
	\begin{tabular}{ccccccc} 
		JDB &   RV  &   RV error    &   FWHM    &   BIS &   Exptime &   Instrument\\
		(-2400000)	& (\kms)& (\kms) &(\kms) & (\kms) &  (s)   &  \\
		\hline
58111.8084	&	35.4749	&	0.0369	&	11.7616	&	-0.2434	&	2700	&	CORALIE\\
58113.7093	&	36.0010	&	0.0411	&	11.9519	&	0.0337	&	2700	&	CORALIE\\
58118.6806	&	36.0550	&	0.0418	&	12.0847	&	0.0198	&	2700	&	CORALIE\\
58129.8234	&	35.5313	&	0.0702	&	11.8283	&	0.0679	&	1800	&	CORALIE\\
58169.7427	&	35.4951	&	0.0448	&	12.0594	&	0.0809	&	2700	&	CORALIE\\
58172.5402	&	36.0431	&	0.0497	&	12.0437	&	-0.0176	&	2700	&	CORALIE\\
58195.5527	&	35.6563	&	0.0450	&	11.8628	&	-0.0829	&	2700	&	CORALIE\\
58201.7069	&	35.6593	&	0.0735	&	12.1195	&	0.0072	&	1800	&	CORALIE\\
58202.7184	&	36.0204	&	0.0382	&	11.9054	&	-0.0995	&	2700	&	CORALIE\\
58207.5924	&	36.0663	&	0.0410	&	11.7546	&	-0.0344	&	2700	&	CORALIE\\
58208.5990	&	35.7431	&	0.0561	&	11.8260	&	-0.0614	&	2700	&	CORALIE\\
58209.5025	&	35.5571	&	0.0513	&	12.0882	&	0.0690	&	2700	&	CORALIE\\
58210.4972	&	35.7017	&	0.0624	&	12.1724	&	0.0087	&	1800	&	CORALIE\\
58211.5079	&	36.0788	&	0.0566	&	11.8893	&	0.2033	&	1800	&	CORALIE\\
58212.5487	&	36.0007	&	0.0556	&	12.0946	&	-0.1138	&	1800	&	CORALIE\\
58214.4983	&	35.5283	&	0.0796	&	12.3596	&	0.1707	&	1800	&	CORALIE\\
		\hline
	\end{tabular}
\end{table*}


\section{Analysis}
\label{sec:analysis}

\subsection{Stellar Properties}
\label{sub:stellar}

\subsubsection{Gaia} \label{secGaia}
\begin{figure}
    \centering
	\includegraphics[width=\columnwidth, angle=0]{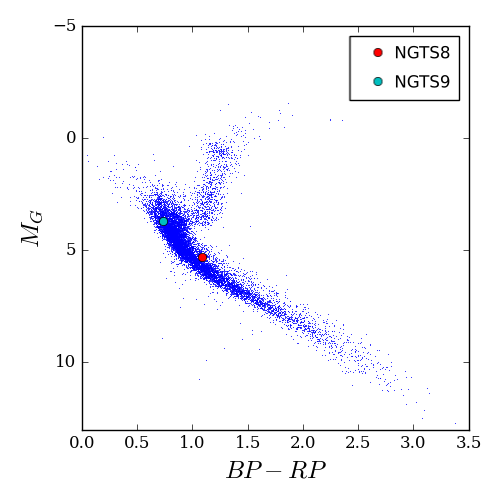}
    \caption{HR diagram using Gaia DR2 absolute magnitude.
    The K0V star, \Nstar, is shown in red and the F8V, \NNstar\ in light blue.}
    \label{fig:gaia_hr}
\end{figure}

To obtain astrometric information for \Nstar\ and \NNstar\ we crossmatched both sources with \gaia\ DR2. 
To check the quality of the astrometric solutions we calculated the unit weight error (UWE) and then renormalised UWE (RUWE). We find that both sources pass the filters recommended by the \gaia\ team \citep[RUWE\,<\,1.4, see][for a discussion on the recommended UWE filters]{Lindegren18}. Along with this, the two targets also have zero astrometric noise, giving us confidence that they are both single sources without evidence of unresolved binarity. They also pass the photometric filters specified by \citet{Arenou2018} to identify blended stars. With the \gaia\ information for each source, we calculate the absolute magnitude and plot their positions on the Hertzsprung-Russell diagram in Figure~\ref{fig:gaia_hr}, where we can see that both \Nstar\ and \NNstar\ lie in the region expected for single main sequence stars.

\subsubsection{\texttt{SPECIES}}
The stacked spectra for both targets were also analyzed using \texttt{SPECIES} \citep{SotoJenkins2018}, a python tool to derive stellar parameters in an automated fashion from high resolution echelle spectra. By measuring the equivalent widths (EWs) for a list of irons lines, and by using the ATLAS9 model atmospheres \citep{ATLAS9}, \texttt{SPECIES} first solves the radiative transfer equation using MOOG \citep{Sneden1973}. From an iterative process, \texttt{SPECIES} derives then the atmospheric parameters (T$_{\rm eff}$, log g, $[$Fe/H$]$) of our target. By interpolation through a grid of MIST isochrones \citep{Dotter2016}, the mass and radius are estimated, using a Bayesian approach. This method delivers an estimate of the age of the system as well. However, due to the fact that solar-type stars spend most of their lives in the evolutionary stage and because the dependence of their effective temperature and luminosity with the age of the system is weak, the estimation of the age of the system is very unconstrained for main sequence stars. Finally, \texttt{SPECIES} derives the rotational and macroturbulent velocities from the stellar temperature and by line-fitting to a set of five absorption lines.
Parameters found by \texttt{SPECIES} for both targets are displayed in Table~\ref{tab:stellar1} and Table~\ref{tab:stellar2}.

\begin{table}
	\centering
	\caption{Stellar Properties for \Nstar.}
	\begin{tabular}{lcc} 
	Property	&	Value		&Source\\
	\hline
    \multicolumn{3}{l}{Astrometric Properties}\\
    R.A.		& \NRA			&2MASS	\\
	Dec			& \NDec			&2MASS	\\
    2MASS I.D.	& \Ntwomass & 2MASS \\
    Gaia source I.D. & \Ngaia & Gaia DR2 \\
    $\mu_{{\rm R.A.}}$ (\masy) & \NpropRA & Gaia DR2 \\
	$\mu_{{\rm Dec.}}$ (\masy) & \NpropDec & Gaia DR2 \\
    parallax (mas)	& \Nparallax		&Gaia DR2	\\    
    \\
    \multicolumn{3}{l}{Photometric Properties}\\
	
	V (mag)		&\NVmag 	&APASS\\
	B (mag)		&\NBmag		&APASS\\
	g (mag)		&\Ngmag		&APASS\\
	r (mag)		&\Nrmag		&APASS\\
	i (mag)		&\Nimag		&APASS\\
    G (mag)		&\NGAIAmag	&Gaia DR2\\
    $\text{G}_{\text{RP}}$ (mag) &\NGAIARPmag	&Gaia DR2\\
    $\text{G}_{\text{BP}}$ (mag) &\NGAIABPmag	&Gaia DR2\\
    J (mag)		&\NJmag		&2MASS	\\
   	H (mag)		&\NHmag		&2MASS	\\
	K (mag)		&\NKmag		&2MASS	\\
    W1 (mag)	&\NWmag		&WISE	\\
    W2 (mag)	&\NWWmag	&WISE	\\
    W3 (mag)	&\NWWWmag	&WISE	\\
    \\
    \multicolumn{3}{l}{Derived Properties}\\
    Spectral type       & K0V       & Gaia DR2 \\
    T$_{\rm eff}$ (K)    & \NteffSPECIES        & SPECIES \\
    $[$Fe/H$]$   & \NmetalSPECIES      & SPECIES \\
    $v \sin i_*$ (\kms)	     & \NvsiniSPECIES       & SPECIES \\
    vmac (\kms)          & \NvmacSPECIES        & SPECIES \\
    log g                & \NloggSPECIES        & SPECIES \\
    \mstar (\msun)       & \NstarmassSPECIES    & SPECIES \\
    \rstar (\rsun)       & \NstarradiusSPECIES  & SPECIES \\
    Age (Gyrs)           & \NstarageSPECIES     & SPECIES \\
    Distance (pc)       & \Ngaiadistance        & Gaia DR2 \\
	\hline
    \multicolumn{3}{l}{2MASS \citep{2MASS}; APASS \citep{APASS};}\\
    \multicolumn{3}{l}{WISE \citep{WISE};}\\
    \multicolumn{3}{l}{Gaia DR2 \citep{gaia_dr2}}\\
	\end{tabular}
    \label{tab:stellar1}
\end{table}

\begin{table}
	\centering
	\caption{Stellar Properties for \NNstar.}
	\begin{tabular}{lcc}
	Property	&	Value		&Source\\
	\hline
    \multicolumn{3}{l}{Astrometric Properties}\\
    R.A.		& \NNRA			&2MASS	\\
	Dec			& \NNDec			&2MASS	\\
    2MASS I.D.	& \NNtwomass & 2MASS \\
    Gaia source I.D. & \NNgaia & Gaia DR2 \\
    $\mu_{{\rm R.A.}}$ (\masy) & \NNpropRA & Gaia DR2 \\
	$\mu_{{\rm Dec.}}$ (\masy) & \NNpropDec & Gaia DR2 \\
    parallax (mas)	& \NNparallax		&Gaia DR2	\\    
    \\
    \multicolumn{3}{l}{Photometric Properties}\\
	V (mag)		&\NNVmag 	&APASS\\
	B (mag)		&\NNBmag		&APASS\\
	g (mag)		&\NNgmag		&APASS\\
	r (mag)		&\NNrmag		&APASS\\
	i (mag)		&\NNimag		&APASS\\
    G (mag)		&\NNGAIAmag	&Gaia DR2\\
    $\text{G}_{\text{RP}}$ (mag) &\NNGAIARPmag	&Gaia DR2\\
    $\text{G}_{\text{BP}}$ (mag) &\NNGAIABPmag	&Gaia DR2\\
    J (mag)		&\NNJmag		&2MASS	\\
   	H (mag)		&\NNHmag		&2MASS	\\
	K (mag)		&\NNKmag		&2MASS	\\
    W1 (mag)	&\NNWmag		&WISE	\\
    W2 (mag)	&\NNWWmag	&WISE	\\
    W3 (mag)	&\NNWWWmag	&WISE	\\
    \\
    \multicolumn{3}{l}{Derived Properties}\\
    Spectral type       & F8V       & Gaia DR2 \\
    T$_{\rm eff}$ (K)    & \NNteffSPECIES        & SPECIES \\
    $[$Fe/H$]$  & \NNmetalSPECIES       & SPECIES \\
    $v \sin i_*$ (\kms)	     & \NNvsiniSPECIES       & SPECIES \\
    vmac (\kms)          & \NNvmacSPECIES        & SPECIES \\
    log g                & \NNloggSPECIES        & SPECIES \\
    \mstar (\msun)       & \NNstarmassSPECIES    & SPECIES \\
    \rstar (\rsun)       & \NNstarradiusSPECIES  & SPECIES \\
    Age (Gyrs)           & \NNstarageSPECIES     & SPECIES \\
    Distance (pc)       & \NNgaiadistance        & Gaia DR2 \\
	\hline
    \multicolumn{3}{l}{2MASS \citep{2MASS}; APASS \citep{APASS};}\\
    \multicolumn{3}{l}{WISE \citep{WISE};}\\
    \multicolumn{3}{l}{Gaia DR2 \citep{gaia_dr2}}\\
	\end{tabular}
    \label{tab:stellar2}
\end{table}

\subsection{Stellar Activity and Rotation on \Nstar}
\label{sub:activity}

In addition to modeling the stellar parameters, we also attempted to search for stellar activity and rotation signals. As mentioned earlier, the out-of-eclipse light curves of both targets show no appreciable variability, and there are no correlations with the measured RVs and the bisector or the FWHM for the two targets -- see Section~\ref{sec:spec}. Nonetheless, determining the stellar rotation period, along with knowledge of the stellar radius and $v \sin i_*$, can enable the inclination angle of the stellar rotation axis to be constrained. This can enable misaligned star-planet systems to be identified \citep{Watson2010}.

\begin{figure}
	\includegraphics[width=\columnwidth, angle=0]{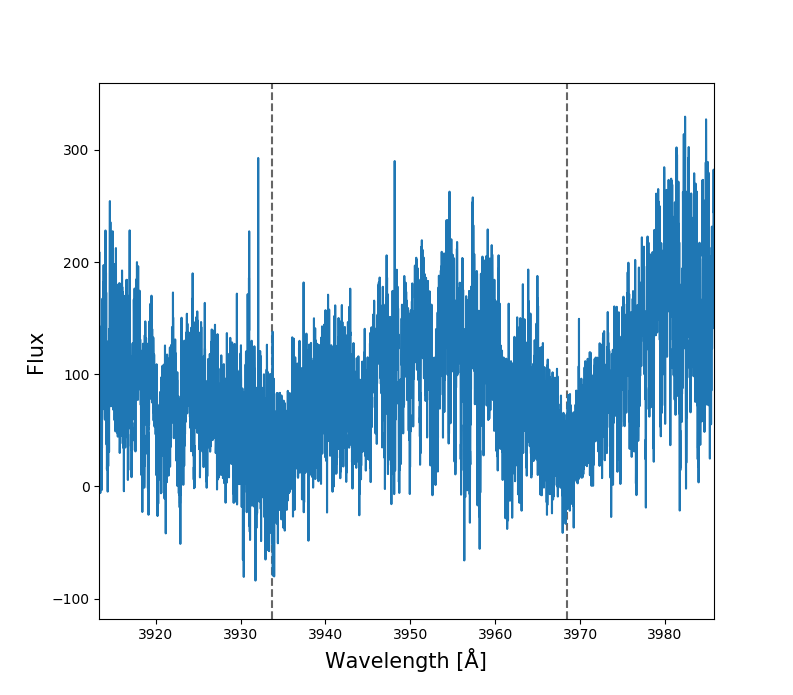}
    \caption{Stacked spectrum of \Nstar\ from HARPS, zoomed on the H (3933.664 \AA) and K (3968.470 \AA) bands, represented with dashed lines.}
    \label{fig:ActivityK0V}
\end{figure}

In order to put constraints on the stellar rotation period, two methods were used. The first one consists of using the activity of the star. Using the formulae described in \cite{Lovis2011}, the $\log R'_{HK}$ was measured for each individual HARPS spectrum. Since the $\log R'_{HK}$ of \NNstar, an F8V star, was not measurable, we will only focus on the K0V star, \Nstar, in this section. The calculated value of the $\log R'_{HK}$ varied between $-4.643$ and $-5.053$ in the individual spectra due to a low signal-to-noise ratio in the blue band, with a mean value of $-4.783$. In order to increase the precision of the measured data, a stacked spectrum of all the spectra was created. Figure~\ref{fig:ActivityK0V} shows this spectrum, zoomed on the H (3933.664\,\AA) and K (3968.470\,\AA) bands, represented with dashed lines. Using this spectrum, we found a value of $-4.817 \pm 0.110$ for the $\log R'_{HK}$. Finally, using the relation from \cite{Noyes1984}, the stellar rotation period of \Nstar\ was derived to be $37.7 \pm 4.1$ days.

Our second approach used the $v \sin i_*$ of the star measured from \texttt{SPECIES} and assumed that the stellar inclination angle, $i_*$, was 90$^{\circ}$ (i.e.  $\sin(i_*) = 1$). This then enables an upper limit to be placed on the stellar rotation period when the stellar radius is known (e.g. \citealt{Watson2010}). An upper limit of $13.92 \pm 2.64$ days was found, which is discrepant with our previous result by almost 5 sigma and would rule out the long rotation period inferred from the $\log R'_{HK}$ measurement. Even adopting the most extreme individual $\log R'_{HK}$ measurement implies a stellar rotation period greater than 26 days.

Given this discrepancy, we decided to verify the $v \sin i_*$ value measured by SPECIES with another technique. We did this by taking a stellar spectrum of a slowly rotating star of the same spectral type as \Nstar\, and artificially broadening it by different $v \sin i_*$ amounts (using a Gray rotational broadening profile). The projected rotational broadening of \Nstar\, was then measured
using an optimal-subtraction technique in which the broadened template spectra were multiplied by a constant and then subtracted
from the \Nstar\, spectrum. This is done after correcting for
radial velocity shifts and re-interpolating to a constant velocity scale. The value of the rotational broadening is then the one that minimises the scatter in the residual spectrum after performing the optimal subtraction. For our template spectrum, we used $\alpha$ Cen B, which has a spectral type very close to \Nstar\, and a low rotation rate. We constructed the template spectrum by stacking archival HARPS spectra taken over 1 night when $\alpha$ Cen B was known to be inactive. The result of this analysis yielded a value consistent with that found by SPECIES.

Adopting the firm upper-limit on the stellar rotation period from the $v \sin i_*$ measurement would ordinarily lead to a much higher $\log R'_{HK}$ level than the one observed. While no definitive answer can explain this difference, some systems hosting hot-Jupiters are known to have suppressed Ca\,II\,H\,\&\,K re-emission (e.g. WASP-12 -- \citealt{Fossati2013}), leading to a lower measured value of the $\log R'_{HK}$. However, these systems generally contain hot-Jupiters very close to filling their Roche lobes, which is not the case for our planet, \Nplanet. We therefore conclude that the most likely explanation of the discrepancy is that we have caught \Nstar\ in an extended low-activity state.

\subsection{Global Modelling}
\label{sub:global}

Analysis of the different photometric and spectroscopic data was performed on \Nstar\ and \NNstar\ data using \texttt{allesfitter} \citep[][and in prep.]{allesfitter-ascl}.
\texttt{allesfitter} is a user-friendly and publicly available software package for modeling data from photometric and RV instruments. Its generative model can account for multi-star systems, stellar flares, star spots and multiple exoplanets.
For this, it constructs an inference framework that unites the versatile packages
\texttt{ellc} \citep[light curve and RV models;][]{Maxted2016}, 
\texttt{aflare} \citep[flare model;][]{Davenport2014},
\texttt{dynesty} \citep[nested samplingl;][]{Speagle2019},
\texttt{emcee} \citep[MCMC sampling;][]{Foreman-Mackey2013}, and 
\texttt{celerite} \citep[GP models;][]{Foreman-Mackey2017}.
\texttt{allesfitter} is accesible at \url{https://github.com/MNGuenther/allesfitter}.

For \Nstar\ and \NNstar, the Nested Sampling approach (see \citealt{skilling2004}) was used, which enables simultaneous fitting of the transit light curves and radial velocity data. 
In particular, we fit for the following astrophysical parameters: a planet's orbital period $P$, the transit epoch $T_C$, the radius ratio $R_p/R_\star$, the sum of radii over the semi-major axis $(R_p + R_\star)/a$, the cosine of the inclination $\cos{i}$, the eccentricity and argument of periastron parameterized as $\sqrt{e} \sin{\omega}$ and $\sqrt{e} \cos{\omega}$, and the RV semi-amplitude $K$.
For the transit light curve modeling, a quadratic limb-darkening law was adopted parameterized after \citet{Kipping2013} as $u_1$ and $u_2$.
Systematic trends in the transit light curves were modeled by a Gaussian process with Matern 3/2 kernels parameterized by the GP's amplitude $\ln{\rho}$ and time scale $\ln{\sigma}$.
For both planets, all photometric data were used for the fits as well as all spectroscopic data, with instrumental offsets taken into account, relevant for \Nstar\, where HARPS and FEROS data were combined for the modelling.

We find that \Nplanet\ has a mass of \Nmass\,\mjup\ and a radius \Nradius\,\rjup, while \NNplanet\ has a mass of \NNmass\,\mjup\ and a radius \NNradius\,\rjup.
The results of the fits for the two planets are summarized in Tables~\ref{tab:planet1} and~\ref{tab:planet2} and shown in earlier plots. Figure~\ref{fig:ngtsphot1} shows (in red) 20 light curve models generated from randomly drawn posterior samples of the \texttt{allesfitter} fit to the NGTS, SAAO and Euler light curves, respectively, for \Nstar. In the same way, Figure~\ref{fig:ngtsphot2} shows the photometric data of NGTS, SAAO and Euler, respectively, for \NNstar, with (in red) 20 light curve models generated from randomly drawn posterior samples of the \texttt{allesfitter} fit. For the RV data, Figure~\ref{fig:harps1} shows the modelling of HARPS,  blue points, and FEROS, orange points, for \Nstar\ and  Figure~\ref{fig:harps2} shows the modelling of the CORALIE data for \NNstar.

In order to check our results, we also performed another analysis of the photometric data from NGTS and available spectroscopic data on \Nstar\ and \NNstar\ using the EXOplanet traNsits and rAdIal veLocity fittER (\texttt{EXO-NAILER} -- \citealt{Espinoza2016}). Using the Markov chain Monte Carlo (MCMC) with a total of 250 walkers for 20000 jumps and 5000 burn-in steps, the modelling was done assuming pure white-noise for the inputted light curves. A logarithmic limb-darkening law was adopted with limb-darkening coefficients taken from \cite{Claret2013}, and sampled according to \cite{Espinoza2015}. The results from this second analysis all agreed, within the error bars, with those found from \texttt{allesfitter}.

\subsection{TESS}
\label{sub:TESS}

\begin{figure}
	\centering
	\includegraphics[width=\columnwidth]{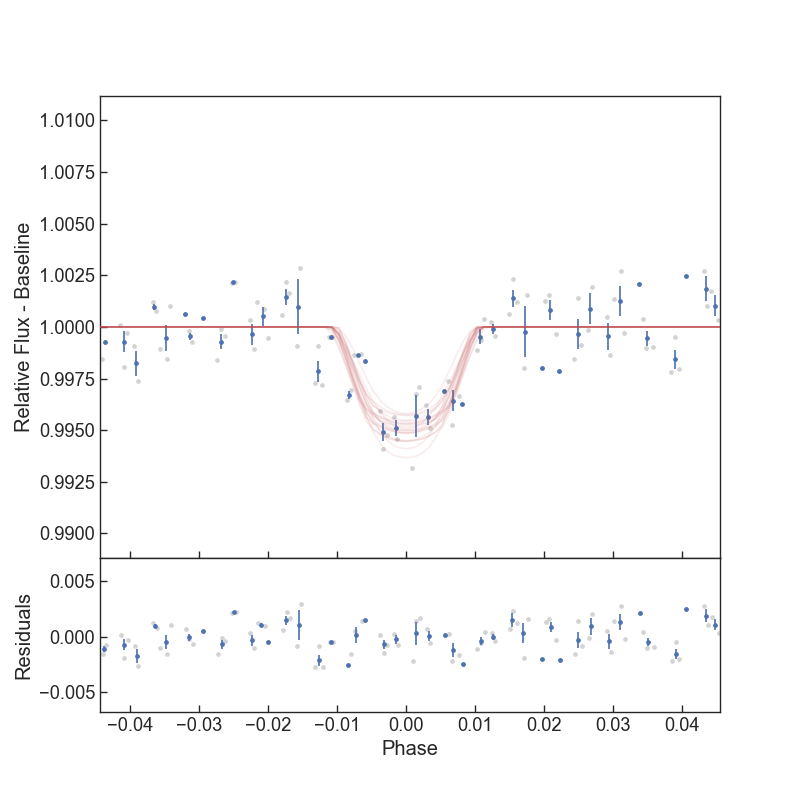}
   \caption{TESS light curve of \NNplanet\ with residuals. The blue data points are binned every 7\,min to aid visualisation. The red lines show 20 light curve models generated from randomly drawn posterior samples of the \texttt{allesfitter} fit.}
    \label{fig:TESSphot1}
\end{figure}

During the preparation of this manuscript TESS photometry was released for \NNstar, which was observed in Sector 8. In response to this data release, we re-analysed, using all available data, \NNstar\ with \texttt{allesfitter}. The TESS photometric data is presented in Figure~\ref{fig:TESSphot1} with the models generated from the fit. We confirmed that using TESS data in our modelling of \NNstar\ did not change or improve the values obtained, and thus the TESS data was not taken into account in the analysis presented in this work. The fact that TESS does not improve the results can be explained by the magnitude of \NNstar, $V = 12.80 \pm 0.02$. At these magnitudes we have found that NGTS and TESS perform similarly \citep{Wheatley2018}. No TESS data is available for \Nstar.

\begin{table}
	\centering
	\caption{Planetary properties for \Nplanet\ using \texttt{allesfitter}.}
	\begin{tabular}{lc} 
	Property	&	Value \\
	\hline
    P (days)		&	\Nperiod	\\
	T$_C$ (BJD)		&	\Ntc	\\
    T$_{14}$ (hours) & \Nduration\\
    $a/R_{*}$		& \Naoverr\\
    $R/R_{*}$       & \Nrratio\\
	K (\ms) 	&\Nkamp	\\
    e 			& \Necc  	\\
    i (degrees)  & \Ninc \\
    \mpl (\mjup)& \Nmass	\\
    \rpl (\rjup)& \Nradius  \\
    $\rho_{p}$ (\gccc) & \Ndensity\\
    a (AU) & \Nau \\
    T$_{eq}$ (K) & \NTeq	\\
	\hline
	\end{tabular}
    \label{tab:planet1}
\end{table}

\begin{table}
	\centering
	\caption{Planetary properties for \NNplanet\ using \texttt{allesfitter}.}
	\begin{tabular}{lc}
	Property	&	Value \\
	\hline
    P (days)		&	\NNperiod	\\
	T$_C$ (BJD)		&	\NNtc	\\
    T$_{14}$ (hours) & \NNduration \\
    $a/R_{*}$		& \NNaoverr \\
    $R/R_{*}$       & \NNrratio\\
	K (\ms) 	&\NNkamp	\\
    e 			& \NNecc  	\\
    i (degrees)  & \NNinc \\
    \mpl (\mjup)& \NNmass	\\
    \rpl (\rjup)& \NNradius  \\
    $\rho_{p}$ (\gccc) & \NNdensity\\
    a (AU) & \NNau \\
    T$_{eq}$ (K) & \NNTeq	\\
	\hline
	\end{tabular}
    \label{tab:planet2}
\end{table}

\section{Discussion}
\label{sec:Discussion}

As outlined in the introduction, at incident fluxes greater than $2 \times 10^5$\,W\,m$^{-2}$ (\citealt{Miller2011}; \citealt{Demory2011}), hot-Jupiters are increasingly found with radii that are significantly larger than theoretically predicted (\citealt{Anderson2011}; \citealt{Delrez2016}; \citealt{Almenara2015}). Using Gaia DR2 measurements for the stellar luminosity and the orbital parameters listed in Tables~\ref{tab:stellar1} and~\ref{tab:planet1} for \Nplanet\ and listed in Tables~\ref{tab:stellar2} and~\ref{tab:planet2} for \NNplanet, we calculated the flux received by both planets to be greater than this limit ($6.85 \pm 0.45 \times 10^5$\,W\,m$^{-2}$ and $9.92 \pm 1.09 \times 10^5$\,W\,m$^{-2}$ for \Nplanet\ and \NNplanet,
respectively). Thus, the stellar irradiation levels received by both of these planets puts them firmly in the regime where we might expect them to exhibit larger than predicted planetary radii. 

\cite{Sestovic2018} conducted a statistical investigation on hot-Jupiter radii and found that above a threshold in incident flux ($2 \times 10^5$\,W\,m$^{-2}$ \citealt{Miller2011}; \citealt{Demory2011}), the observed radius follow the thermal evolution models (\citealt{Miller2011}; \citealt{Thorngren2016}) with the addition of an inflation parameter, $\Delta R$. This observed radius `inflation' is dependent on both the incident stellar flux and the mass of the planet. They proposed a flux-mass-radius relationship that has distinct forms for 4 different planetary mass regimes: below 0.37\,\mjup, between 0.37 -- 0.98\,\mjup, between 0.98 -- 2.50\,\mjup\ and over 2.50\,\mjup. Using these relationships, we calculated what would be the expected radius inflation ($\Delta R$) values for our two planets. For \Nplanet, its mass lies on the edge of two regimes in \cite{Sestovic2018} (\mpl\,<\,0.98\,\mjup\ and \mpl\,>\,0.98\,\mjup), we thus determined predicted $\Delta R$s from both relationships of $0.24 \pm 0.02$ and $0.02 \pm 0.01$ \rjup, respectively. While the first value would suggest a highly inflated radius, the second value however suggests almost no inflation. Concerning \NNplanet, the predicted radius inflation, $\Delta R$, is $0.18 \pm 0.01$ \rjup. Thus, from the work of \cite{Sestovic2018}, these two planets would be expected to exhibit planetary radii larger than predicted.

\begin{table*}
	\centering
	\caption{A summary of the mass-radius model for \Nplanet\ with radius \Nradius\,\rjup.}
	\label{tab:modelsummary1}
	\begin{tabular}{cccccccc}
    Model & Mass of the & Orbital &  Age of the & 
    Mass fraction of & Core & Radius\\
        & planet (\mjup) & separation (AU) & system (Gyrs) & heavy material & mass (\%) & (\rjup) \\
	\hline
    \cite{Baraffe2008} & 1 & 0.045 & 8.93 - 10.00 & 0.02 - 0.1 & & 1.025 - 1.074 \\
    \cite{Fortney2007} & 1 & 0.045 & 4.5 & & 0 - 25 & 1.050 - 1.107 \\
    \hline
	\end{tabular}
\end{table*}

\begin{table*}
	\centering
	\caption{A summary of the mass-radius model for \NNplanet\ with radius \NNradius\,\rjup.}
	\label{tab:modelsummary2}
	\begin{tabular}{cccccccc}
    Model & Mass of the & Orbital &  Age of the & 
    Mass fraction of & Core & Radius\\
        & planet (\mjup) & separation (AU) & system (Gyrs) & heavy material & mass (\%) & (\rjup) \\
	\hline
    \cite{Baraffe2008} & 2 - 5 & 0.045 & 0.30 - 1.78 & 0.02 - 0.1 & & 1.063 - 1.177 \\
    \cite{Fortney2007} & 2.44 & 0.045 & 0.30 - 1 &  & 0 - 100 & 1.065 - 1.199 \\
    \hline
	\end{tabular}
\end{table*}

We finally compared the observed planetary radii of \Nplanet\ and \NNplanet\ to the mass-radius models of \cite{Baraffe2008} and \cite{Fortney2007} who present, assuming a solar-type star, tables of planetary radii as a function of core mass, mass of the planet, orbital separation and age of the system. Since neither of the host stars of the planets presented here are solar-like, we had to renormalise the orbital separation in order to keep the same incident flux. In this scenario, the distance from their host star would be equal to $0.044$\,AU and $0.038$\,AU for \Nplanet\ and \NNplanet, respectively. As one can see in Tables~\ref{tab:modelsummary1} and~\ref{tab:modelsummary2}, both models seem consistent and correctly predict the measured radius of \Nplanet, \Nradius\,\rjup, and \NNplanet, \NNradius\,\rjup, using the described parameters.

To conclude, even if both planets are in a regime where we expect planets to exhibit larger than predicted radii, our two planets are non-inflated hot-Jupiters. This could be due to the planets being enriched with heavy elements, yielding a more compact structure and thus a smaller radius, like HD\,149026b \citep{Sato2005}.


\begin{figure}
	\includegraphics[width=\columnwidth, angle=0]{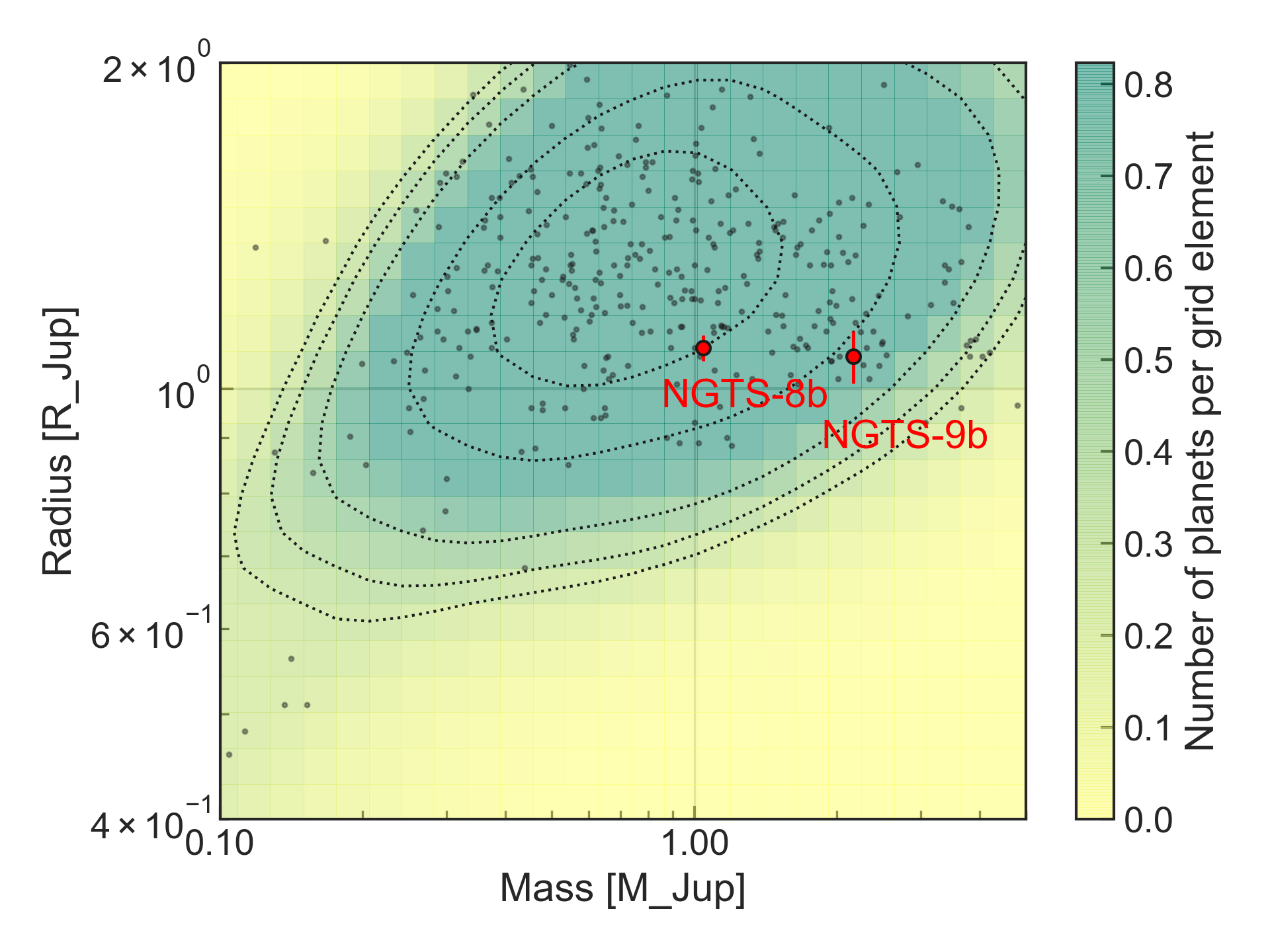}
    \caption{Shows \Nplanet\ and \NNplanet\ in the planetary radius vs planetary mass plot, in regards with all exoplanets from the NASA Exoplanet Archive with radius uncertainty below 10\% or mass uncertainty below 50\% and with an incident flux received by the planets greater than $2 \times 10^5$\,W\,m$^{-2}$. The background and the dotted black lines represents the point density per grid element.}
    \label{fig:periodmass}
\end{figure}


\section{Conclusions}
\label{sec:conclusions}
  
We have presented the latest discovery by the Next Generation Transit Survey (NGTS) of two non-inflated hot-Jupiters: \Nplanet\ and \NNplanet. NGTS, SAAO and Euler photometric data and spectroscopic data from HARPS, FEROS and CORALIE were used to confirm the detection of these two planets. By combining some of these data, an analysis of the transiting planets was performed using \texttt{allesfitter} and confirmed with \texttt{EXO-NAILER}. From this model, both planets have orbits consistent with being circular, as expected for such short period hot-Jupiters. The characteristic of the planets were calculated such as: \Nplanet\ with a mass of \Nmass\,\mjup\ and a radius \Nradius\,\rjup, and \NNplanet\ with a mass of \NNmass\,\mjup\ and a radius of \NNradius\,\rjup. Figure~\ref{fig:periodmass} shows these discoveries in comparison to known planets with radius higher than 0.4\,\rjup. 

A study of the rotational period of the K0V star, \Nstar, was performed using different models and despite a significant discrepancy that we assume is due to an extended low activity of the star, we measured an upper limit of $13.92 \pm 2.64$ days. 
While its host is considerably fainter, our analysis also suggests that its planet, \Nplanet, could have similar properties to HD 189733b, one of the best studied hot-Jupiters. Further observations of \Nplanet\ will allow direct comparisons to be drawn between these two hot-Jupiters. The upcoming launch of JWST will enable high-precision observations of \Nplanet's full-phase curve, which would be of particular interest due to the efficient dayside to nightside heat recirculation that HD 189733b exhibits relative to other hot-Jupiters (\citealt{Knutson2007}; \citealt{Schwartz2017}).

Concerning \NNplanet, the planet is highly irradiated, with an incident flux around $9.59 \pm 0.74 \times 10^5$\,W\,m$^{-2}$, yet non-inflated. This radius could be due to the planet being extremely enriched with heavy elements, explaining its density, \NNdensity\,\gccc, one of the highest compare to planets with similar masses, as shown in Figure~\ref{fig:periodmass}.

\section*{Acknowledgements}
Based on data collected under the NGTS project at the ESO La Silla Paranal Observatory.  The NGTS facility is operated by the consortium institutes with support from the UK Science and Technology Facilities Council (STFC)  project ST/M001962/1.
This paper uses observations made at the South African Astronomical Observatory (SAAO). We thank Marissa Kotze (SAAO) for developing the SHOC camera data reduction pipeline.  
The contributions at the University of Warwick by PJW, RGW, DLP, DJA, BTG and TL have been supported by STFC through consolidated grants ST/L000733/1 and ST/P000495/1. 
Contributions at the University of Geneva by DB, FB, BC, LM, and SU were carried out within the framework of the National Centre for Competence in Research ``PlanetS" supported by the Swiss National Science Foundation (SNSF).
The contributions at the University of Leicester by MRG and MRB have been supported by STFC through consolidated grant ST/N000757/1.
CAW acknowledges support from the STFC grant ST/P000312/1.
EG gratefully acknowledges support from Winton Philanthropies in the form of a Winton Exoplanet Fellowship. 
JSJ acknowledges support by Fondecyt grant 1161218 and partial support by CATA-Basal (PB06, CONICYT).
DJA gratefully acknowledges support from the STFC via an Ernest Rutherford Fellowship (ST/R00384X/1).
PE and HR acknowledge the support of the DFG priority program SPP 1992 "Exploring the Diversity of Extrasolar Planets" (RA 714/13-1).
LD acknowledges support from the Gruber Foundation Fellowship.
MNG acknowledges support from MIT's Kavli Institute as a Torres postdoctoral fellow.
The research leading to these results has received funding from the European Research Council under the FP/2007-2013 ERC Grant Agreement number 336480 and from the ARC grant for Concerted Research Actions, financed by the Wallonia-Brussels Federation. This work was also partially supported by a grant from the Simons Foundation (PI Queloz, ID 327127).
This work has made use of data from the European Space Agency (ESA)
mission {\it Gaia} (\url{https://www.cosmos.esa.int/gaia}), processed by
the {\it Gaia} Data Processing and Analysis Consortium (DPAC,
\url{https://www.cosmos.esa.int/web/gaia/dpac/consortium}). Funding
for the DPAC has been provided by national institutions, in particular
the institutions participating in the {\it Gaia} Multilateral Agreement.
PyRAF is a product of the Space Telescope Science Institute, which is operated by AURA for NASA.
This research has made use of the NASA Exoplanet Archive, which is operated by the California Institute of Technology, under contract with the National Aeronautics and Space Administration under the Exoplanet Exploration Program.




\bibliographystyle{mnras}
\bibliography{paper} 








\bsp	
\label{lastpage}
\end{document}